**In Situ Observations of Preferential Pickup Ion Heating at an Interplanetary Shock**

E. J. Zirnstein[1*], D. J. McComas[1,2], R. Kumar[1], H. A. Elliott[2], J. R. Szalay[1], C. B. Olkin[3], J. Spencer[3], S. A. Stern[3], L. A. Young[3]

[1]Department of Astrophysical Sciences, Princeton University, Princeton, NJ 08544, USA
[2]Southwest Research Institute, San Antonio, TX 78238, USA
[3]Southwest Research Institute, Boulder, CO 80302, USA

Abstract:

Non-thermal pickup ions (PUIs) are created in the solar wind (SW) by charge-exchange between SW ions (SWIs) and slow interstellar neutral atoms. It has long been theorized, but not directly observed, that PUIs should be preferentially heated at quasi-perpendicular shocks compared to thermal SWIs. We present in situ observations of interstellar hydrogen ($H^+$) PUIs at an interplanetary shock by the *New Horizons'* Solar Wind Around Pluto (SWAP) instrument at ~34 au from the Sun. At this shock, $H^+$ PUIs are only a few percent of the total proton density but contain most of the internal particle pressure. A gradual reduction in SW flow speed and simultaneous heating of $H^+$ SWIs is observed ahead of the shock, suggesting an upstream energetic particle pressure gradient. $H^+$ SWIs lose ~85% of their energy flux across the shock and $H^+$ PUIs are preferentially heated. Moreover, a PUI tail is observed downstream of the shock, such that the energy flux of all $H^+$ PUIs is approximately six times that of $H^+$ SWIs. We find that $H^+$ PUIs, including their suprathermal tail, contain almost half of the total downstream energy flux in the shock frame.

## I. Introduction

As the solar wind (SW) expands outward from the Sun into interplanetary space, slow interstellar neutral atoms (mostly hydrogen, H) flowing into the heliosphere interact with SW ions (SWIs) via charge-exchange [1]. The ionized interstellar neutral atoms are "picked up" by the motional electric field of the SW, hence their name pickup ions (PUIs). During the pickup process, newly-injected PUIs first form a narrow ring beam in velocity space and then subsequently scatter onto an isotropic shell distribution. The Coulomb collisional time for protons is significantly larger than the SW propagation time, thus, PUIs do not thermalize with the SWIs [2]. Interstellar PUIs have been observed by, e.g., *Ulysses* SWICS out to ~5 au [3], revealing a high acceleration efficiency for PUIs at interplanetary shocks [4], though SWIs still contain the majority of the plasma pressure at this distance and dominate the shock interaction.

*New Horizons'* Solar Wind Around Pluto (SWAP) [5] instrument utilizes a top-hat electrostatic analyzer to detect ions in the energy range ~0.021-7.8 keV/q [6]. It has made high resolution measurements of the SW out to ~41 au from the Sun [6,7]. SWAP also uses its large field of view to provide high quality measurements of PUI speed distributions. McComas et al. [7] provided the first analysis of interstellar PUIs co-moving with the SW out to ~38 au from the Sun, quantifying the PUI density, temperature, and internal pressure from SWAP measurements and extrapolating their moments to the SW termination shock (TS), offering key predictions for outer heliosphere studies. Since H$^+$ PUIs dominate the internal plasma pressure beyond ~20 au [7], it is believed that they should have a significant effect on the energy dissipation at interplanetary shocks. It has been theorized [8-10] and inferred from *Voyager 2* in situ measurements [11] that non-thermal PUIs should be preferentially heated at quasi-perpendicular shocks compared to thermal ions; however, this has not yet been observed.

In this Letter, we provide the first in situ observations of the preferential heating of H$^+$ PUIs at an interplanetary shock. We analyze a particular shock that was observed by SWAP at ~34 au from the Sun when both H$^+$ SWIs and H$^+$ PUIs were measured. This shock is intriguing because the interaction appears quite similar to *Voyager 2* observations at the TS [11], although *Voyager 2* was unable to observe PUIs. Observations of a PUI-mediated shock provides important insights into other shocks in the heliosphere. For example, observations show that there is a significant suprathermal particle population in the inner heliosheath downstream of the TS [12,13]. These populations are important for understanding, for example, the plasma pressure gradients in the heliosheath [14] as well as their contribution to energetic neutral atoms observed at 1 au by NASA's *Interstellar Boundary Explorer* [15-17].

## II. Observations and Analysis

At approximately 02:11 UTC on 2015 October 5, the SWAP instrument aboard *New Horizons* observed an interplanetary shock with a ~17% jump in SW speed from ~380 to 440 km s$^{-1}$, and a significant increase in H$^+$ SWI temperature (~1100%) downstream of the shock (Fig. 1). While the cadence of SWAP measurements of SWIs is ~10 minutes, interstellar H$^+$ PUIs are measured using 1-day histograms of SWAP count rates to compute more accurate moments of the PUI distribution [7]. Nonetheless, we are able to determine that the average PUI filled-shell density increased by a factor of ~2.5 and temperature increased by ~65% across of the shock.

We estimate the shock speed, $V$, in the Sun frame using the change in H$^+$ PUI density from upstream ($n_1$) to downstream ($n_2$) of the shock, such that $V = (n_2 u_2 - n_1 u_1)/(n_2 - n_1)$ where $u$ is the SW speed in the Sun frame. We use the PUI density, rather than the SWI density, to compute the shock strength since it appears that the SWI density fluctuates due to other SW disturbances unrelated to the shock, while the PUIs remain stable for several days before and after the shock. In fact, if we compute the 1-day average SWI density before and after the shock at the same time scale as the PUIs, the SWI density actually decreases by ~10%. Note that, as we show later, a fraction of PUIs form a suprathermal tail downstream of the shock. The PUI tail is also included in the calculation of the compression ratio.

We find that the density compression ratio $n_2/n_1 = 3.0$ and shock speed $V = 475$ km s$^{-1}$. This compression ratio is slightly higher than that observed by *Voyager 2* at the TS [11]. The *Voyagers* were not able to directly measure PUIs in the SW or at the TS. However, SWAP observations show that PUIs already dominate the internal pressure in the SW by ~20 au from the Sun, with an ever-increasing number density fraction with distance, so that they surely contain the vast majority of internal pressure at the TS [7]. Thus, we provide a comparison between SWAP and *Voyager 2* observations in Fig. 2 to better understand the role of PUIs at heliospheric shocks. A comparison between their measurements upstream and downstream of the shocks is shown in the Supplementary Material [18].

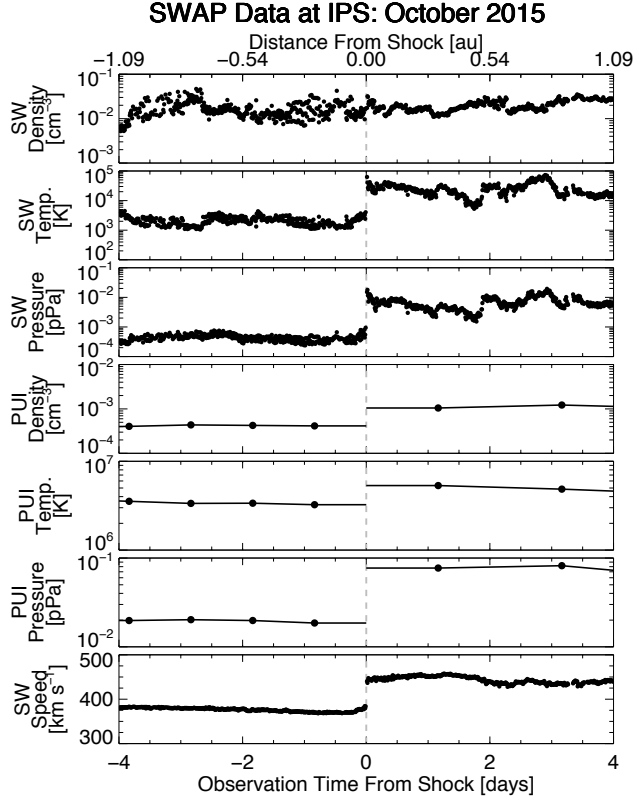

**FIG 1.** SWAP observations at an interplanetary shock (IPS) in October 2015. The top *x*-axis labels show the distance from the shock derived in the shock frame. Since the PUI data cadence is ~1 day, we connect the data points with lines and plot horizontal lines from the two PUI data points nearest to the shock. Note that there is no PUI data ~2 days after the shock due to culling [7].

An interesting aspect of the SWAP observations is that there is a gradual reduction of the SW speed by ~10% (in the shock frame) within ~0.07 au ahead of the shock (Fig. 2). There is a corresponding increase in H$^+$ SWI temperature by ~100%, likely a result of adiabatic compression of the slowing SW plasma. A distance of ~0.07 au is much larger than the H$^+$ PUI gyro-radius (~10$^5$ km or ~10$^{-4}$ au, for 0.1 nT magnetic field), suggesting that this is created by a positive gradient in high energy (~MeV) particle pressure ahead of the shock [19,20]. The decrease in dynamic pressure by the slowing of the SW gives an estimate of the energetic particle pressure at the shock front, yielding ~0.03 pPa or ~0.2 eV cm$^{-3}$. This behavior is similar to the ~15% slowing observed by *Voyager 2* starting ~0.7 au ahead of the TS, which inferred ~0.1 eV cm$^{-3}$ energetic particle pressure [21].

At ~34 au from the Sun, PUIs are only a few percent of the proton number density [18,22], and thus produce an internal pressure much smaller compared to the SW dynamic pressure. At the TS, the PUI density is expected to be ~15-30% of the total density [7,15], such that the PUI internal pressure is ~10-20% of the SW dynamic pressure. Nevertheless, PUIs gain a significant fraction of energy across the interplanetary shock observed by SWAP despite their low number density. To quantify this, we calculate the energy density flux $E_i$ (hereafter "energy flux") for each particle species (subscript *i*),

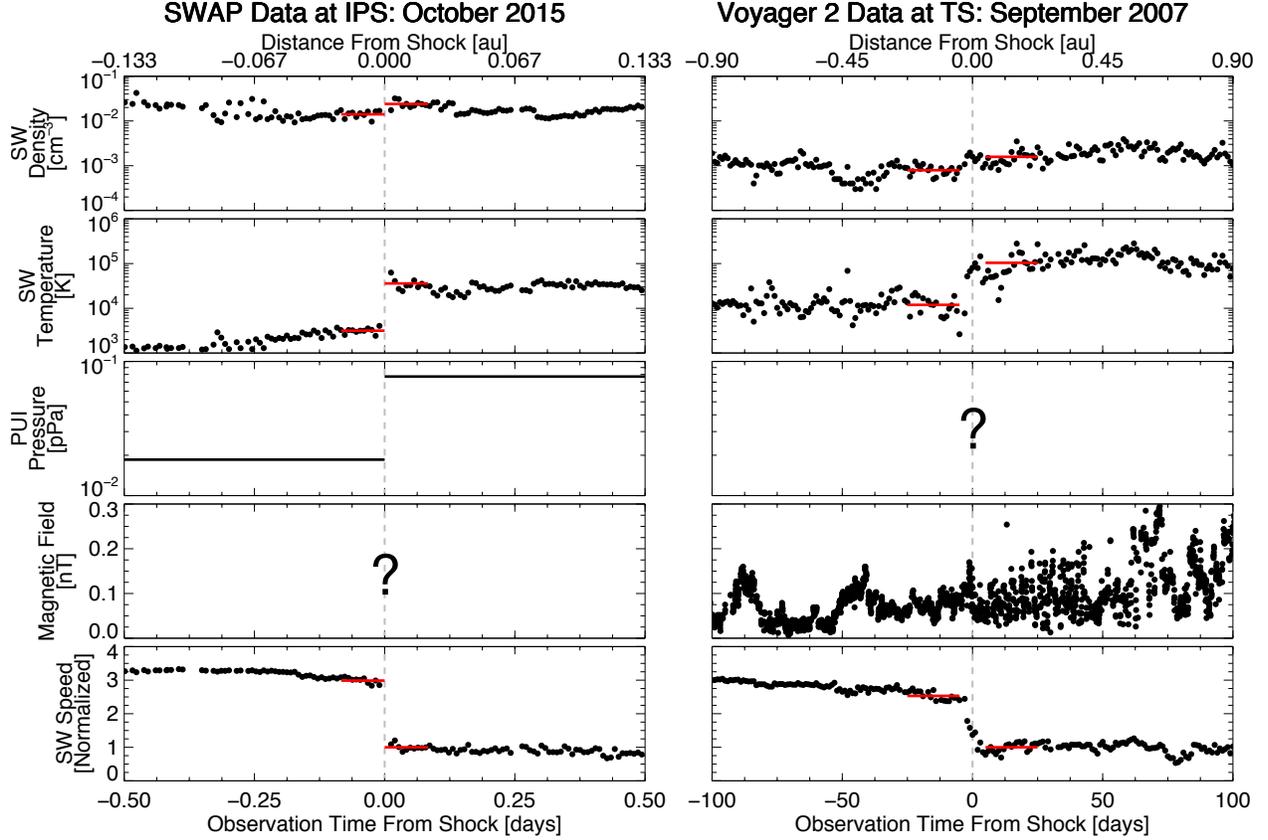

**FIG 2.** (left) SWAP observations at the interplanetary shock. The middle of the SWAP PUI measurement times are outside the *x*-axis range, thus we show horizontal lines at their levels before and after the shock. (right) *Voyager 2* observations at the TS. We show daily-averaged particle moments and hourly-averaged magnetic field. Red lines indicate the average before and after the shocks for SWAP and *Voyager 2* data shown in the Supplement [18], except for the magnetic field. SW speeds are transformed to the shock frame (*V* - *u*), then normalized to the average downstream value indicated by the red line.

$$E_i = \frac{1}{2} m_i n_i u_S^3 + \frac{\gamma}{\gamma-1} n_i k_B T_i u_S, \quad (1)$$

where $n_i$ is number density, $T_i$ is temperature, $m_i$ is mass, $\gamma = 5/3$ is the adiabatic index, $k_B$ is Boltzmann's constant, and $u_s$ is the SW bulk flow speed in the shock frame. Eq. (1) is derived from the magnetohydrodynamic energy conservation equation across a perpendicular shock [18]. The density and temperature of each species are computed from the integration of the particle distributions derived from the fitting analysis.

The particle energy flux is shown in Fig. 3a. Since the H$^+$ PUI measurements are made every ~24 hours, we linearly interpolate the H$^+$ PUI data to the resolution of the H$^+$ SW data. For the two PUI data points nearest to the shock, we assume the PUI density and temperature are constant up to the shock jump. We only show data for H$^+$ SWIs and H$^+$ PUIs in Fig. 3a. Below, we discuss the contributions of electrons, alphas (He$^{++}$), other non-thermal particles, and the magnetic field to the total energy flux. Note that the small-scale fluctuations seen in the PUI energy flux in Figure 3, as well as the steady decline in PUI energy flux within ~0.25 days ahead of the shock, are due to changes in the SW bulk flow speed in the shock frame, $u_S$, in Eq. 1.

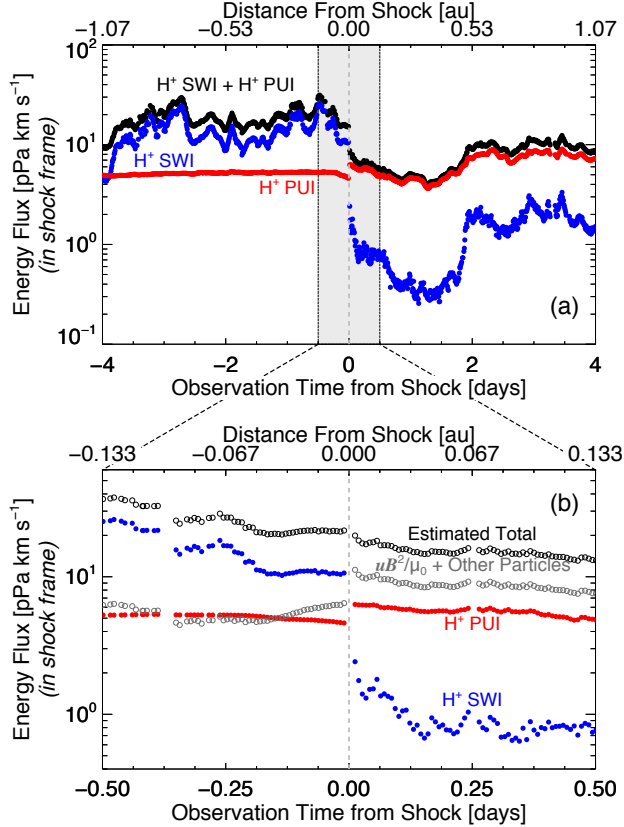

**FIG 3.** (a) Energy flux for H$^+$ SWIs (blue), H$^+$ PUIs (red), and their total (black) in the shock frame. We perform one-hour boxcar smoothing over SW density and speed. PUI data are interpolated to the SWI measurement resolution. (b) Energy flux close to the shock. We also show the estimated energy flux contribution from the magnetic field, alphas, He$^+$ PUIs, H$^+$ PUI tail, electrons, and energetic particles (gray open circles), and the estimated total (black open circles). Note that the PUI density and temperature are assumed constant in panel b using the two PUI data points closest to the shock (horizontal lines in Figure 1 right before and after the shock).

    The total energy flux (particles plus magnetic field) should be conserved across the shock. However, the energy flux of each particle species will change depending on their interaction with the shock. In Fig. 3a, H$^+$ SWIs have ~70% of the total observed energy flux (H$^+$ SWIs plus H$^+$ PUIs) upstream of the shock, while H$^+$ PUIs have ~30%. H$^+$ SWIs lose ~85% of their energy flux across the shock and H$^+$ PUIs increase by ~30%. The decrease in SW energy flux, which is strikingly similar to what *Voyager 2* observed at the TS (note that we show energy density flux, and Richardson et al. [11] show energy per particle), and the preferential heating of PUIs across the shock is clear evidence that non-thermal particles, including PUIs, modify the shock structure [23]. Downstream of the shock, the H$^+$ PUI energy flux is approximately four times that of H$^+$ SWIs. Note, however, that while the majority of the upstream energy flux is contained in H$^+$ SWIs and PUIs, their combined energy flux downstream is smaller than that upstream by ~50%. This difference is significantly larger than the expected change in magnetic energy flux across the shock [18], indicating that we are not accounting for all of the particles.

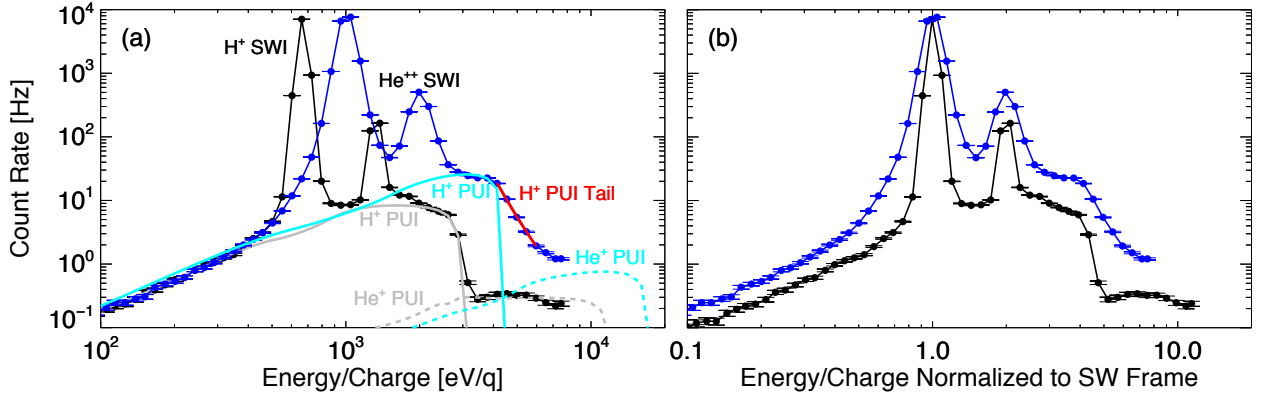

**FIG 4.** (a) SWAP day-averaged count rates in the spacecraft frame before (black) and after (blue) the shock (see Fig. 1). Fits to the H$^+$ PUIs before and after the shock are shown in gray and cyan, respectively. A fit to the H$^+$ PUI tail after the shock is shown in red. Models of the He$^+$ PUIs are shown as dashed lines. (b) Data are normalized to the SW frame.

Interestingly, SWAP count rates show a tail at energies above the H$^+$ PUI cutoff downstream of the shock (Fig. 4). Before the shock, the H$^+$ SWIs (peaked at ~650 eV/q in Fig. 4a, or 1 in Fig. 4b) and alphas (twice the energy/charge) are relatively cold, and the H$^+$ PUI distribution is well represented by a filled-shell function with cutoff at approximately twice the SW speed. After the shock, H$^+$ SWIs, alphas, and H$^+$ PUIs are all hotter and denser (the count rates increase and broaden in energy), but there is also a tail population at energies above the H$^+$ PUI cutoff which was not included in the H$^+$ PUI filled-shell fit [7].

We compute the H$^+$ PUI tail energy flux by fitting a power-law speed distribution in the SW frame to the 5 energy bins above the H$^+$ PUI filled-shell cutoff (before He$^+$ PUIs) after converting to SWAP count rates. We determine the best-fit function to be $f(v) = 1134$ [s$^3$ km$^{-6}$] $(v/u_c)^{-9.7}$, where $v$ is the particle speed and $u_c$ is the H$^+$ PUI filled-shell cutoff speed, both in the SW frame. Due to the very steep slope, the majority of the PUI tail density is within the fitted energy range. The H$^+$ PUI tail density is ~$1.9 \times 10^{-4}$ cm$^{-3}$, approximately 15% of the total downstream H$^+$ PUI density, and the effective temperature is ~$1.1 \times 10^7$ K. Based on these derived parameters, it appears possible that the PUI tail originated from H$^+$ PUIs that were energized at the shock by, for example, reflection from the cross-shock potential and energization in the upstream motional electric field [9,23,24]. The steepness of the PUI tail appears reasonable under this scenario since this is not likely diffusive shock acceleration or particle interactions with turbulence, which would likely result in a harder spectrum. Interestingly, the PUI tail persists for ~2-3 days downstream of the shock, where the spectral slope slightly softens before the tail disappears.

While SWAP does not measure the magnetic field or electrons, and it is difficult to quantify the alpha and He$^+$ PUI distributions directly from SWAP observations, we can estimate their contribution to the total energy flux. First, we determine the electron density assuming the plasma is quasi-neutral, and that electrons have the same temperature as H$^+$ SWIs upstream and downstream of the shock. This assumption is reasonable based on theoretical arguments of electron temperatures in the SW [25]. Though some electrons may accelerate to non-thermal energies at the shock, it is unlikely they hold a significant fraction of the downstream pressure [23]. Second, we assume the alpha number density is 4% of H$^+$ SWIs (based on SW data extracted from OMNIWeb at 1 au ~4-5 months earlier) and their temperature is 4.5 times that of H$^+$ SWIs based on their collision-less nature [26]. We note that our results are not sensitive to assumptions for the alpha particles due to their low number density.

Next, we calculate the He$^+$ PUI distribution upstream of the shock [7] using the Vasyliunas & Siscoe [27] distribution and scale the density to match the He$^+$ PUI shelf (~4-8 keV/q in Fig. 4a). To estimate the He$^+$ PUI distribution downstream, following Zank et al. [9,24] we assume that the majority of He$^+$ PUIs increase in temperature similarly to the H$^+$ PUIs (temperature increased by ~65%), but a fraction of them (proportional to $\sqrt{Zm_H/m_{He}}$ = 0.5 times the reflection efficiency of H$^+$ PUIs (15%), or 0.5×15% = 7.5%) may be further energized at the shock with a temperature increasing by a factor of $(m_{He}/m_H)^2$ = 16 times greater than H$^+$ PUIs. Then, we include the high energy particle pressure gradient ahead of the shock calculated above, assuming it increases linearly with distance starting from 0.07 au upstream of the shock and reaches 0.03 pPa at the shock front, with a constant pressure downstream. Finally, we include the magnetic field energy flux. In lieu of in-situ magnetic field measurements, as *New Horizons* is not equipped with a magnetometer, we assume that the magnetic field magnitude upstream of the shock is equal to the median value measured by *Voyager 2* from ~22 to 39 au from the Sun (0.15 nT) [18,28].

Including these populations in the total energy flux, as well as the H$^+$ PUI tail downstream of the shock, yields a nearly constant energy flux across the shock (Fig. 3b). While our calculation of the total energy flux has uncertainties from, e.g., estimates of the magnetic field and measurement errors [18], our analysis strongly indicates that H$^+$ PUIs hold a significant fraction of the total downstream energy flux. Considering the possible range of magnetic field magnitude [18], H$^+$ PUIs hold between ~30% and ~60% of the downstream energy flux, while H$^+$ SWIs are only ~5-10%. The remaining downstream energy flux is in the magnetic field, alphas, He$^+$ PUIs, electrons, and high energy particles combined. Thus, this study provides the first direct observation of the mediation and preferential heating of non-thermal PUIs, rather than the thermal SWIs, at a shock, where PUIs (including the tail) hold approximately half of the total downstream energy flux.

*Acknowledgements.* E.Z. acknowledges support from NASA grant NNX16AG83G. This work was also carried out with partial support from the *IBEX* mission, which is part of NASA's Explorer Program. D.M. and H.E. acknowledge support from the SWAP instrument effort on the *New Horizons* mission, which is part of NASA's New Frontiers Program. H.E. also acknowledges support from NASA grant NNX12AB26G. R.K. acknowledges support from the Max-Planck/Princeton Center for Plasma Physics and NSF grant AST-1517638. The authors thank Kimberly Ennico and Hal Weaver for their roles as Project Scientists for the *New Horizons*' mission and Randy Gladstone for leading the *New Horizons*' Particles and Plasma Theme Team. We acknowledge the use of *Voyager 2* plasma data published online by the MIT Space Plasma Group: http://web.mit.edu/space/www/voyager.html, and *Voyager 2* magnetic field data from OMNIWeb: https://omniweb.gsfc.nasa.gov. SWAP H$^+$ SWI and H$^+$ PUI data are publicly available online at CDAWeb: https://cdaweb.sci.gsfc.nasa.gov/index.html.

*Corresponding author: ejz@princeton.edu

**Supplementary Material:**

**TABLE SI.** Comparison of SWAP measurements at the interplanetary shock and *Voyager 2* measurements at the termination shock (TS).

| Parameter | SWAP [a] | | *Voyager 2* [a] | |
|---|---|---|---|---|
| | Upstream | Downstream | Upstream | Downstream |
| Flow speed [km s$^{-1}$] [b] | 96 | 32 | 326 | 129 |
| H$^+$ SW density [cm$^{-3}$] | 1.4×10$^{-2}$ | 2.4×10$^{-2}$ | 7.9×10$^{-4}$ | 1.6×10$^{-3}$ |
| H$^+$ PUI shell density [cm$^{-3}$] | 4.1×10$^{-4}$ | 1.0×10$^{-3}$ | - | - |
| H$^+$ PUI tail density [cm$^{-3}$] | - | 1.9×10$^{-4}$ | | |
| H$^+$ SWI temperature [K] | 3.2×10$^3$ | 3.6×10$^4$ | 1.2×10$^4$ | 1.0×10$^5$ |
| H$^+$ PUI shell temperature [K] | 3.2×10$^6$ | 5.3×10$^6$ | - | - |
| H$^+$ PUI tail temperature [K] | - | 1.1×10$^7$ | | |
| H$^+$ SWI pressure ($nk_BT$) [pPa] | 6.2×10$^{-4}$ | 1.2×10$^{-2}$ | 1.3×10$^{-4}$ | 2.3×10$^{-3}$ |
| H$^+$ PUI shell pressure ($nk_BT$) [pPa] | 1.9×10$^{-2}$ | 7.7×10$^{-2}$ | - | - |
| H$^+$ PUI tail pressure ($nk_BT$) [pPa] | - | 2.9×10$^{-2}$ | | |
| Magnetic field [nT] [c] | 0.15 | 0.45 | 0.05-1 | - |
| Sonic Mach number [d] | 2.8 | 0.48 | 25 | - |
| Alfvén Mach number [d] | 3.8 | 0.56 | 4-8 | - |
| Compression ratio [e] | 3.0 | | 2.5 | |

[a]SWAP and *Voyager 2* measurements are averaged over the intervals shown in Fig. 2 of the main text.
[b]In the shock frame. We assume the TS is stationary.
[c]For *Voyager 2*, we report the range of values ahead of the TS. For SWAP, we assume the upstream magnetic field is the median value measured by *Voyager 2* measurements between ~25 and 39 au from the Sun [1].
[d]The adiabatic index γ = 5/3. Mach numbers for SWAP include H$^+$ SWIs, H$^+$ PUIs, and estimates for alphas and electrons (see main text). *Voyager 2* only includes H$^+$ SWIs.
[e]Ratio of upstream-to-downstream flow speed.

**Rankine-Hugoniot Jump Conditions**

The magnetohydrodynamic jump conditions for a perpendicular shock are

$$[\rho u_S] = 0, \tag{S1}$$

$$\left[\rho u_S^2 + P + \frac{B^2}{2\mu_0}\right] = 0, \tag{S2}$$

$$[u_S B] = 0, \tag{S3}$$

$$\left[\frac{\rho u_S^3}{2} + \frac{\gamma u_S P}{\gamma - 1} + \frac{u_S B^2}{\mu_0}\right] = 0, \tag{S4}$$

where [ ] denotes the change across the shock, $\rho$ is mass density, $u_S = V - u$ is the SW flow speed in the shock frame, $P$ is the internal particle pressure, $B$ is the magnetic field, $\gamma = 5/3$ is the adiabatic index, and $\mu_0$ is the magnetic constant. We utilize Eq. (S4) to calculate the energy flux of the particles and magnetic field, which should be approximately constant across the shock.

**PUI Measurement Uncertainties**

There are both systematic and statistical uncertainties in the analysis of the H+ PUI data. Fitting the spectra with the Vasyliunas & Siscoe [2] function to the PUIs introduces a systematic uncertainty because we are not able to fit to the entire PUI energy range. As described by McComas et al. [3], we fit to (1) energy steps less than one half of the energy/charge of the highest data point in the SW peak and (2) the three energy/charge steps just below the PUI cutoff. We employ these limitations in order to avoid background from internal scattering in the instrument from the SW beam, as well as the potential addition of SWIs and alphas into the PUI fit distribution. While not anticipated to be large, the inability to measure PUIs under the SWI and alpha peaks adds an unquantifiable systematic uncertainty to our calculation of the PUI distribution.

We can, however, quantify the statistical uncertainties from fitting to the H$^+$ PUI measurements (both the filled-shell and tail). We do this by propagating the 1-sigma uncertainties of the fitting functions' parameters from the Vasyliunas & Siscoe function ($\lambda$, $\beta$) for the PUI shell (see McComas et al. [3] for more details on Eq. S5 parameters):

$$f = \frac{3N\beta_E r_E^2 \Theta(1-w)}{8\pi u_p \tilde{u}^3 r w^{3/2}} exp\left(-\frac{\lambda}{rw^{3/2}}\frac{\theta}{\sin\theta}\right), \tag{S5}$$

as well as the parameter uncertainties from the power-law function for the PUI tail ($A$, $\gamma$):

$$f = A\left(\frac{v}{u_c}\right)^{-\gamma}. \tag{S6}$$

To compute the uncertainties of the PUI density ($n$) and temperature ($T$), we first individually add the 1-sigma uncertainties to the free parameters ($\sigma_\lambda$, $\sigma_\beta$, $\sigma_A$, $\sigma_\gamma$):

$$\lambda \to \lambda + \sigma_\lambda,$$
$$\beta \to \beta + \sigma_\beta,$$
$$A \to A + \sigma_A,$$
$$\gamma \to \gamma + \sigma_\gamma, \tag{S7}$$

and calculate the "perturbed" density ($\tilde{n}$) and temperature ($\tilde{T}$) from each individual term in Eq. (S7) in the zeroth and second moment integrals of Eqs. (S5) and (S6). Then, we calculate the difference between the perturbed solution and the original solution due to each parameter and add these deviations in quadrature (assuming they are independent random errors). Thus, the propagated uncertainties for the PUI shell are:

$$\sigma_{n,s} = \sqrt{(n_s - \tilde{n}_s^\lambda)^2 + \left(n_s - \tilde{n}_s^\beta\right)^2},$$

$$\sigma_{T,s} = \sqrt{(T_s - \tilde{T}_s^\lambda)^2 + \left(T_s - \tilde{T}_s^\beta\right)^2}. \tag{S8}$$

The propagated uncertainties for the PUI tail are:

$$\sigma_{n,t} = \sqrt{(n_t - \tilde{n}_t^A)^2 + (n_t - \tilde{n}_t^\gamma)^2},$$

$$\sigma_{T,t} = \sqrt{(T_t - \tilde{T}_t^A)^2 + (T_t - \tilde{T}_t^\gamma)^2}. \tag{S9}$$

For the two PUI data points closest to the shock upstream and downstream, we find that the relative 1-sigma uncertainties of the PUI shell moments upstream to be 3.5% (density) and 0.9% (temperature) and downstream to be 2.3% (density) and 0.4% (temperature). The relative uncertainties in the PUI tail downstream are calculated to be 1.2% and <0.1% for density and temperature, respectively. The statistical uncertainties are significantly smaller than the relative energy fluxes of the PUIs compared to the total energy flux (between ~30% and 60%, see below).

**Uncertainties in SWI and PUI Energy Fluxes Due to Magnetic Field**

For our calculations of the total energy flux, we assume the magnetic field upstream of the shock is equal to the median value observed by *Voyager 2* at similar distances from the Sun [1]. Using this value (0.15 nT), we estimate that $H^+$ PUIs hold ~50% of the total downstream energy flux. Bagenal et al. [1] determine that the 10th and 90th percentiles for the magnetic field magnitude observed by *Voyager 2* are 0.08 and 0.28 nT, respectively. Using these values, we estimate that $H^+$ PUIs hold between ~30% ($B = 0.28$ nT) and ~60% ($B = 0.08$ nT) of the downstream energy flux, $H^+$ SWIs hold between ~5% ($B = 0.28$ nT) and ~10% ($B = 0.08$ nT), and the magnetic field between ~60% ($B = 0.28$ nT) and ~10% ($B = 0.08$ nT). Thus, despite the lack of magnetic field measurements from *New Horizons*, $H^+$ PUIs hold a significant fraction of the total downstream energy flux compared to $H^+$ SWIs.